\documentclass[twocolumn,showpacs,preprintnumbers,amsmath,amssymb]{revtex4}
\usepackage{graphicx}

\begin{document}

\preprint{APS/123-QED}

\title{Controlling transitions in a Duffing oscillator by sweeping parameters in time.}

\author{Oleg Kogan}
\email{oleg@caltech.edu} \affiliation{Caltech, Pasadena CA}

\date{\today}
\begin{abstract}
We consider a high-$Q$ Duffing oscillator in a weakly non-linear
regime with the driving frequency $\sigma$ varying in time between
$\sigma_i$ and $\sigma_f$ at a characteristic rate $r$. We found
that the frequency sweep can cause controlled transitions between
two stable states of the system. Moreover, these transitions are
accomplished via a transient that lingers for a long time around
the third, unstable fixed point of saddle type. We propose a
simple explanation for this phenomenon and find the transient
life-time to scale as $-(\ln{|r-r_c|})/\lambda_r$ where $r_c$ is
the critical rate necessary to induce a transition and $\lambda_r$
is the repulsive eigenvalue of the saddle.  Experimental
implications are mentioned.
\end{abstract}
 \maketitle 

Even the simplest of non-linear dynamical systems,
such as those described by second order ODEs are notorious for
exhibiting rich phenomenology~\cite{Strogatz}.  One of the
features of such systems is multi-stability, which strictly exists
when all parameters are time-independent. In the case of
adiabatically varying parameters, a system initially situated at
one of the quasi-fixed points will remain close to it.  What
happens when the parameters vary faster then the time scales
determined by eigenvalues around the fixed points (FP) will be
explored here on the case of a damped, driven Duffing oscillator.
When the variation of parameters is sufficiently rapid,
transitions occur, such that after the rapid variation is
finished, the system finds itself at the FP different from the one
at which it was initially situated.

\begin{figure}[h]
\begin{center}
\includegraphics[width=8cm]{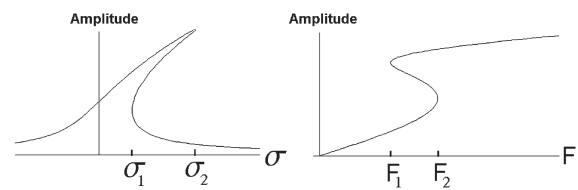}
\end{center}
\caption{\label{fig:Fig1}Left: constant $F$.  Right: constant
$\sigma$.}
\end{figure}

\begin{figure}[h]
\begin{center}
\includegraphics[width=8cm]{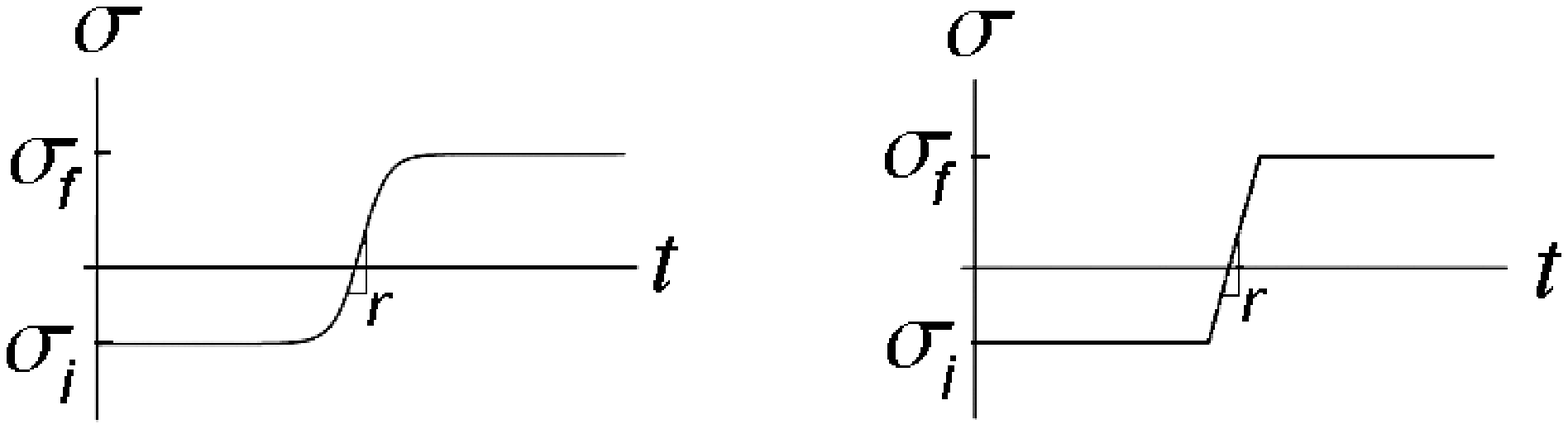}
\end{center}
\caption{\label{fig:Fig2}$\sigma(t)$.}
\end{figure}

\emph{Statement of the Problem.} We will consider an
oscillator obeying the following equation of motion:
\begin{eqnarray}
\label{eq:Eq1} 2f \cos{\phi(t)} &=& \ddot{x} + 2\lambda\dot{x} +
\omega_0^2
x\left(1+\alpha x^2\right)\\
\dot{\phi} &=& \gamma(t) \nonumber
\end{eqnarray}
where $\omega_0$ is the angular frequency of infinitesimal
vibrations, $\lambda$ is the damping coefficient, which for a
quality factor $Q$ is given by $\lambda = \frac{\omega_0}{2Q}$,
$\alpha$ is a non-linear coefficient, $f$ is the driving strength,
and $\gamma$ is the driving frequency.  Upon non-dimensionalizing
and re-scaling we obtain the following equation which has the form
of a perturbed simple harmonic oscillator:
\begin{eqnarray}
\label{eq:Eq2} \ddot{x} + x &=& \epsilon\left(2F\cos{\phi(t)} -
\dot{x} -
x^3\right) \\
\dot{\phi} &=& 1 + \epsilon \sigma(t) \nonumber
\end{eqnarray}
For the case of time-independent parameters $F$ and $\sigma$ the
amplitude response has the well-known frequency-pulled form
~\cite{LL, Nayfeh}.  For $F > F_{cr}$ there is a region of
tri-valuedness with two stable branches and one unstable (middle)
branch (Fig.~\ref{fig:Fig1}).  Such response curves have recently
been measured for NEMS~\cite{Ali}, indicating their Duffing-like
behavior. For a constant $\sigma$ but different $F$ there is also
a response function with tri-valuedness for $\sigma
> \sigma_{cr}$. To each case there corresponds also a tri-valued
phase response (not shown).
\\
We would like to explore what happens if the driving frequency,
starting from the single-valued region (at $\sigma_i$) is rapidly
varied in time, into the tri-valued region (ending at $\sigma_f$),
at a constant F. We will also explore the phenomenology resulting
from varying $F$ at a constant $\sigma$ in a similar manner - from
a single-valued regime into the tri-valued regime.  The variation
will take place via a function with a step - either smooth (such
as $\tanh$) or piece-wise linear - Fig.~\ref{fig:Fig2}.
In this article we will explore the situation when $F/F_{cr} \sim
O(1)$, for instance $F = 2 F_{cr}$. When $F \sim F_{cr}$, the size
of the hysteresis $(\sigma_2 - \sigma_1)$ is also of order $1$ and
the perturbation method is especially simple. The more complicated
case of $F \gg F_{cr}$ (but small enough that r.h.s. of
Eq.~(\ref{eq:Eq2}) is still a perturbation over harmonic
oscillator equation) may be considered in a future work.

\emph{Phenomenology.}
We discover the following: upon sweeping $\sigma$ from lower
values into the hysteresis, depending on other parameters, the
solution of Eq.~(\ref{eq:Eq2}) may jump unto the lower branch. The
transitions have a peculiar feature of having lifetimes much
longer then the slow time scale of Eq.~(\ref{eq:Eq2}) - the
damping time-scale of order $\epsilon$.
\begin{figure}[h]
\begin{center}
\includegraphics[width=8cm]{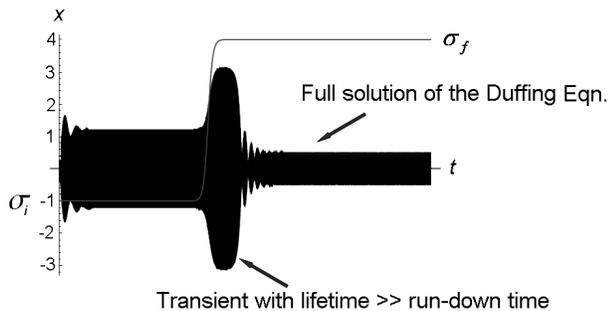}
\end{center}
\caption{\label{fig:Fig3}Example of a transition.}
\end{figure}
\\
Now, there are four relevant "control knobs": $F$, $\sigma_i$,
$\sigma_f$ and $r$ defined here as $1/($sweeping time$)$. Consider
a set of imaginary experiments, each experiment performed for a
different sweep rate $r$, with all three other parameters fixed.
As $r$ gets larger, depending on the value of other parameters
there may be a critical sweep rate $r_{cr}$ beyond which
transitions will be induced. Moreover for $r$ approaching very
close to $r_{cr}$ the life-time of transitions, $\tau$, will grow.
We will sample these lifetimes and $r-\sigma_f$ configuration
space numerically and explain the finding theoretically below.
However one more observation, depicted in Fig.~\ref{fig:Fig4} is
in order. For $r$ close to $r_{cr}$, not only do the life-times,
$\tau$ of transients tend to grow, but also the amplitudes and
phases of these long transients tend to approach that of the
middle (unstable) branch of the static Duffing oscillator.
\begin{figure}[h]
\begin{center}
\includegraphics[width=7.2cm]{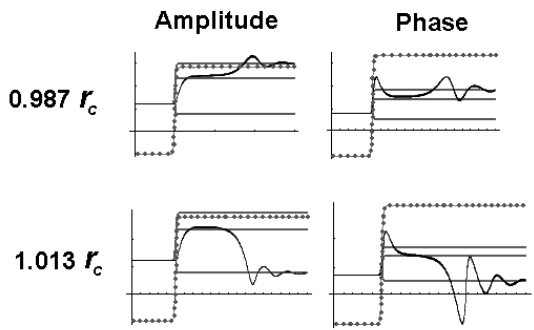}
\end{center}
\caption{\label{fig:Fig4} Solutions kiss the unstable branch -
envelope of the full numerical solution of Eq.~(\ref{eq:Eq1})
along with quasi fixed points (solutions to
Eq.~(\ref{eq:AmpEq1})-(\ref{eq:AmpEq2}) below with lhs set to $0$)
are shown. The function $\sigma(t)$ is displayed as a dotted
curve.}
\end{figure}
\\ We learn that for $r$ close $r_{cr}$ the solution moves unto the unstable
branch, lives there for a time period $\tau$ and then either moves
unto the top branch if $r<r_{cr}$ or performs the transition unto
the bottom branch if $r>r_{cr}$.  The jump unto either the top or
the bottom branch takes place \emph{long} after reaching the
static conditions (see Fig.~\ref{fig:Fig3}).
A numerical experiment performed at particular parameter values
(see Fig.~\ref{fig:Fig5}) demonstrates the typical situation:
plotted on the semi-log plot, the life-time versus $|r-r_c|$
nearly follows a straight line.  Thus, we learn that $\tau \propto
- \ln\left|r-r_c \right|$.
\begin{figure}[h]
\begin{center}
\includegraphics[width=8cm]{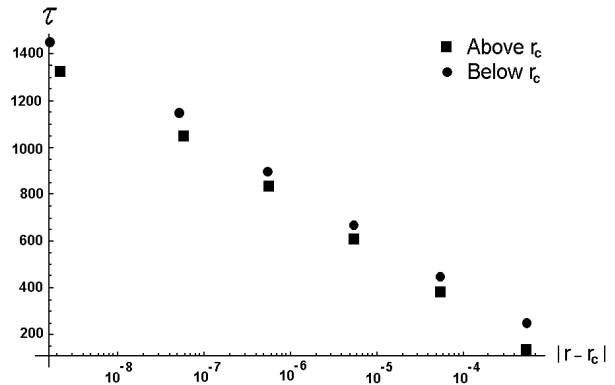}
\end{center}
\caption{\label{fig:Fig5} Transient lifetime versus $|r-r_{cr}|$
for $F\approx 2.8 F_{cr}$, $\sigma_i = -2$, $\sigma_f = 0.5
(\sigma_1 + \sigma_2)$}
\end{figure}
\\
Another numerical experiment was performed to measure $r_c$ versus
$\sigma_f$ for $F \approx 2.8 F_{cr}$ and $\sigma_i = -2$ (see
Fig.~\ref{fig:Fig6}).
\begin{figure}[h]
\begin{center}
\includegraphics[width=9cm]{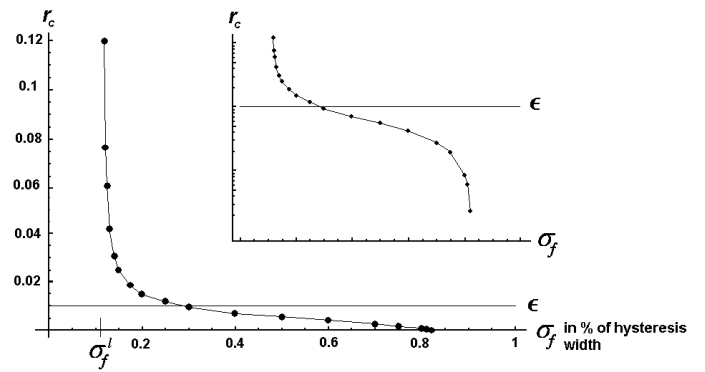}
\end{center}
\caption{\label{fig:Fig6} $F = 2.8 F_{cr}$, $\sigma_i=-2$.  There
is a singularity at $\sigma^l_f$. The size of the hysteresis was
calculated with the first-order theory and therefore it is an
over-estimation - the point where $r_c$ goes to zero (i.e. at
infinitessimal sweep rates) is the exact right edge of the
hysteretic region.  The existence of singularity at $\sigma^l_f$
is a genuine effect, since the first-order theory predicts the
lower end of the hysteresis more accurately.}
\end{figure}
\\
Two interesting features are immediately observed:  (i) the
critical rate necessary to induce a transition is not simply
determined by the small parameter $\epsilon$ in the
Eq.~(\ref{eq:Eq2}); (ii) when sweeping not deeply enough into the
tri-valued region no transitions can be induced for \emph{any}
value of the sweep rate.  Increasing $\sigma_i$ moves the
singularity of $r(\sigma_f)$ to larger values and varying $F$ does
not significantly affect the shape of the $r(\sigma_f)$ curve.  It
is somewhat surprising to see singularities in this rather simple
problem, especially before the onset chaotic regime.
Transitions may occur by sweeping $\sigma$ down; also by sweeping
$F$ and holding $\sigma$ constant.  Transition phenomenon persists
for $F$ large enough that the hysteresis is large (see note in
\emph{Discussion} below).

\emph{Theory for small $\Delta \sigma$ case.}
The regime of $F/F_{cr} \sim O(1)$ guarantees that $(\sigma_2 -
\sigma_1) \sim O(1)$, thus if $\sigma_i$ is chosen sufficiently
close to the hysteretic region, the jump in frequency during the
sweep, $\Delta \sigma$ is also $\sim O(1)$.  This paves way for a
simple perturbation method.  For example, we write the
"multiple-scales" perturbative solution to Eq. (2) as $X=
X_0(t,T,...) + \epsilon X_1(t,T,...) + O(\epsilon^2)$ where the
slow time scale $T = \epsilon t$, and in general $T^{(n)} =
\epsilon^n t$.  Plugging this into the equation and collecting
terms of appropriate orders of $\epsilon$ teaches us that $x_0 =
A(T)e^{it} + c.c.$, i.e. the solution is essentially a slowly
modulated harmonic oscillator, with the modulation function
satisfying the following Amplitude Equation (AE):\
\begin{displaymath}
Fe^{i \delta(T)} - 2i\frac{\partial A}{\partial T} - iA -
3\left|A\right|^2A=0
\end{displaymath}
where $\frac{d\delta}{dT} = \epsilon \sigma(t(T))$.  Such AE holds
for any sweep rate as long as $\Delta \sigma \sim O(1)$.  The AE,
broken into real, $x = Re[A]$, and imaginary, $y=Im[A]$, parts
are:
\begin{eqnarray}
\label{eq:AmpEq1} \frac{dx}{dT} &=& -\frac{1}{2}x + \sigma(T)y -
\frac{3}{2}\left(x^2 + y^2\right)y \\
\label{eq:AmpEq2}\frac{dy}{dT} &=& -\sigma(T)x - \frac{1}{2}y +
\frac{3}{2}\left(x^2 + y^2\right)x - \frac{F}{2}
\end{eqnarray}
These AE are well known and appear in similar forms in
literature~\cite{Nayfeh, BM}. For a certain range of parameters
these equations give rise to a two-basin dynamics with a stable
fixed point (FP) inside each basin.  The basins are divided by a
separatrix which happens to be the stable manifold of the unstable
FP of saddle type.
\begin{figure}[h]
\begin{center}
\includegraphics[width=4.7cm]{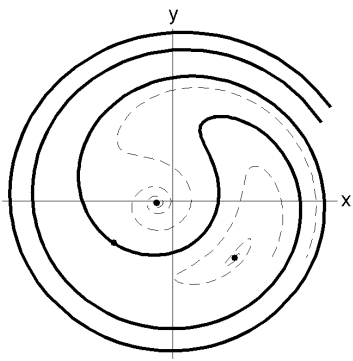}
\end{center}
\caption{\label{fig:Fig7} AE dynamics - an example.  Typical
solutions are shown as dashed curves.  Black dots indicate fixed
points.}
\end{figure}
\\
Such basins have recently have been mapped for a certain type of
NEMS~\cite{BasinMapping}.  Next, consider a set of thought
experiments, each sweeping $\sigma$ at a different sweep rate $r$.
Immediately after the sweep, the system will find itself somewhere
in the two-basin space corresponding to conditions at $\sigma_f$,
call it a point $(x_f(r), y_f(r))$. If $(x_f(r), y_f(r))$ lies in
the basin which has evolved from the basin in which the system
began before the sweep then there will be no transition. If
$(x_f(r), y_f(r))$ lies in the opposite basin, then that
corresponds to a transition! For very low $r$, during the sweep
the system will follow closely to the quasi-FP and there will be
no transition. In the opposite extreme - infinite $r$, at the end
of the sweep the system will not have moved at all, due to
continuity of a dynamical system. This endpoint of
$\{(x_f(\infty), y_f(\infty))\}$ may lie in either basin depending
on the $\sigma_f$.
\begin{figure}[h]
\begin{center}
\includegraphics[width=8cm]{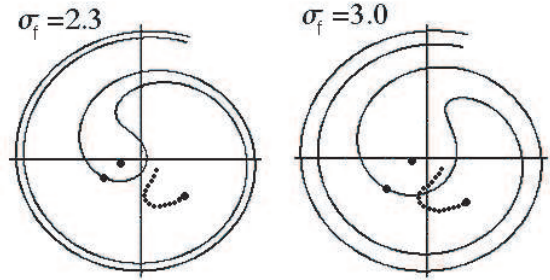}
\end{center}
\caption{\label{fig:Fig8}Real examples at $F = 2.8 F_{cr}$,
$\sigma_i = -2$.  The locus $\{x_f,y_f\}$ is shown by dotted
curves.}
\end{figure}
\\
In the numerical experiments that we considered the first point of
the $\{(x_f(r), y_f(r))\}$ curve to cross the separatrix happens
to be this tail \footnote{We could not prove this to be so for
\emph{any} 2-basin model under \emph{any} sweeping function, but
we expect it for a large class of 2-basin models and sweeping
functions.} at $r=\infty$. This explains the singularity mentioned
earlier \footnote{One can show, using the fact that $r=\infty$
point enjoys the property of being the FP at $\sigma_i$, that the
singularity behaves as $\left(\sigma_f - \sigma_f^l
\right)^{-1/2}$ which agrees well with numerics.} (see
Fig.~\ref{fig:Fig6}).

\emph{Calculation of lifetimes.} The point $(x_f(r), y_f(r))$
serves as an initial condition for subsequent evolution at fixed
$\sigma_f$. The $(x_f(r), y_f(r))$ close to the separatrix (i.e.
for $r$ close to $r_c$) will flow towards the saddle and linger
around it for a while. Because this lingering will take place
close to the saddle, the linearized dynamics around the saddle
should be a good approximation: $\mathbf{r}(T) = \delta l
\mathbf{v}_r e^{\lambda_r T} + R \mathbf{v}_a e^{\lambda_a T}$,
where, for example, $\mathbf{v}_r$ and $\lambda_r$ are a repulsive
eigenvector and eigenvalue, $\delta l$ is the distance of
$(x_f(r), y_f(r))$ away from the separatrix along $\mathbf{v}_r$
and R is the characteristic radius of linearization around the
saddle. The times at which the system crosses this circular
boundary is given by $R = \sqrt{\mathbf{r} \cdot \mathbf{r}}$. The
first time, $T_{in}$, is of the order of $\delta l$.  The second
time is $T_{out} \approx
\frac{1}{\lambda_r}\ln{\left(\frac{R}{\delta l}\right)}$
(neglecting effects of $\mathbf{v}_r$). The lingering time is $\tau =
T_{out} - T_{in} \approx T_{out}$. So,
\begin{equation}
\label{eq:Eq5}
\tau = -\frac{1}{\lambda_r}\ln{\left(\frac{\delta l}{R}\right)} =
-\frac{1}{\lambda_r}\ln{\left(\frac{\frac{dl}{dr}|r-r_c|}{R}\right)}
\end{equation}
Thus we capture the logarithmic dependence of the transient time
versus $|r-r_c|$.
\begin{figure}[h]
\begin{center}
\includegraphics[width=8.7cm]{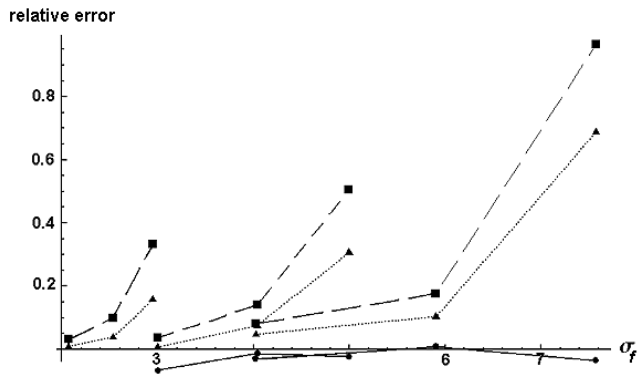}
\end{center}
\caption{\label{fig:Fig9}Dotted lines: Fractional errors between the
slope of $\tau$ vs. $\ln|r-r_{cr}|$ from exact Duffing and
$1/(\epsilon \lambda_r)$ where $\lambda_r$ are computed using the
\emph{first order} perturbation theory in $\epsilon$ (Eqns.
(2)-(4)). Dashed lines: same comparison but with $\lambda_r$
computed to \emph{second order}.  In both cases, $F=1.5$, $F=2$,
$F=2.5$ from left to right. Solid lines: the comparison of
$1/(\epsilon \lambda_r)$ with the same slope from the AE, not full
Duffing (only for $F=2$ and $F=2.5$) - remaining errors may be due
to inexactness of linearized approximation and omission of
$\mathbf{v}_r$.}
\end{figure}
We found that calculating the eigenvalues $\lambda_r$ using the
second order theory as described in~\cite{BM}
\emph{systematically} lowers the discrepancy in the slope of
$\tau$ vs. $|r-r_{cr}|$ computed from exact Duffing and
theoretical predictions, Fig.~\ref{fig:Fig9}, which leads us to
conclude that the errors are due to inexactness of the first order
AE, not due to the incorrectness of the explanation of the cause
of the transition phenomenon.

\emph{Experimental Significance.} One can propose to use the
transition phenomenon with its long transients to position
Duffing-like systems unto the unstable (middle) branch (see
Fig.~\ref{fig:Fig1}) - the desire to do this has been expressed by
workers in the NEMS community~\cite{HenkUnstBr}.  The first
question is whether a necessary $r_c$ is attainable.  We see from
Fig.~\ref{fig:Fig6} and related discussion that for a vast range
of parameters $r_c$ is less then $0.1$. Recall that in this paper
$r$ is defined simply as $1/($sweep time$) = 1/\Delta T$. The more
experimentally relevant quantity is $\Delta \sigma/\Delta T$.  In
the present paper we are concerned with hysteresis widths of order
$1$, so $\Delta \sigma$ in question is $\sim 10$ or less. Hence
$\Delta \sigma/\Delta T$ necessary to create a transition is, for
a vast range of parameters, less then $1$ resonant widths per
run-down time (but close to the high-end of the hysteresis this
figure falls rapidly - see Fig.~\ref{fig:Fig6}), which in
conventional units corresponds to the sweep rate of $(\omega_0/Q)^2$ $Hz/sec$. 
One can also use this as a guide to prevent 
unwanted transitions in experiments in which $\sigma$ depends on
time. The second question is how small must $|r-r_c|$ be to induce
a transient of time $\tau$.  From Eq. (\ref{eqn:Eq5}) we see that $|r-r_c|
\approx \frac{\omega_0}{Q}(dr/dl) \times \exp{\left[-\tau
\lambda_r(\sigma_f, F)\right]}$ sec$^{-1}$.  For any $F$,
$\lambda_r$ reaches maximum approximately in the middle of the
hysteresis where $\lambda_r(\sigma_f = \frac{\sigma_2 +
\sigma_1}{2}, F) \approx 0.64 (F-F_{cr})/F_{cr}$.  So the smaller
the $F$, the easier it is to attain a longer transient. The
quantity $dr/dl$, at the point of crossing the separatrix diverges
at $\sigma_f = \sigma^l_f$ and becomes small ($< 1$) for
$\sigma_f$ close to $\sigma_2$ (this explains why the
life-time $\tau$ is very sensitive to $|r-r_c|$ in this region
(see Eq. (\ref{eq:Eq5})). The functional form of $dr/dl$ versus parameters is
not fundamental - it depends on the form of the ramping function,
for example, and can be computed from the set $\{x_f(r),
y_f(r)\}$.

\emph{Discussion.} Our theoretical hypothesis claims that the
point at which the system finds itself at the end of the ramp,
$(x_f,y_f)$ determines whether there will be a transition or not.
However one may be inquire whether this set of points does not
consistently cross the separatrix in some special way, for
example, always tangentially or always perpendicular to the
separatrix. If it does, there must be something happening during
the ramp that situates these final points in this special way.
Then the theory based on just the final position $\{(x_f,y_f)\}$,
although is true, would be incomplete.  This issue may be
addressed in a future work by analyzing the effect of a generic
perturbation of either the ramping function or the system on
$\{(x_f,y_f)\}$. This might also pave the way to understanding the
generality of the phenomenon - whether it holds in a large class
of two-basin models and ramping functions. Also, we would like to
explore the phenomenology for large $\Delta \sigma$ case when the
AE (\ref{eq:AmpEq1})-(\ref{eq:AmpEq2}) do not hold, yet the system
is still in a weakly non-linear regime.
\\

The author thanks Professor Baruch Meerson for suggesting this
problem and for ideas and time spent in subsequent discussions and
Professor Michael Cross for pointing out the usefulness of
thinking about the set $\{(x_f, y_f)\}$ and its behavior, as well
as for general advice.  We acknowledge the support of the PHYSBIO
program with the funds from the European Union and NATO as well as
the NSF grant award number DMR-0314069.

\end{document}